\documentclass[aps,prd,reprint,nofootinbib,superscriptaddress]{revtex4-2}
\usepackage{amsmath,amssymb,bm}
\usepackage{graphicx}
\usepackage{booktabs}
\usepackage[colorlinks=true,linkcolor=blue,citecolor=blue,urlcolor=blue]{hyperref}
\usepackage[low-sup]{subdepth}

\newcommand{\dd}{\mathrm{d}}
\newcommand{\GeV}{\,\mathrm{GeV}}
\newcommand{\keV}{\,\mathrm{keV}}
\usepackage[low-sup]{subdepth}

\def\Xa{\chi_1^{}}
\def\Xb{\chi_2^{}}

\def\hs{\hspace*{0.3mm}}
\def\hsx{\hspace*{0.5mm}}
\def\hsm{\hspace*{-0.3mm}}

\def\vs{\vspace*{1mm}}

\newcommand{\be}{\begin{equation}}
\newcommand{\ee}{\end{equation}}
\newcommand{\bea}{\begin{eqnarray}}
\newcommand{\eea}{\end{eqnarray}}
\newcommand{\beq}{\begin{equation}}
\newcommand{\eeq}{\end{equation}}
\newcommand{\bq}{\begin{equation}}
\newcommand{\eq}{\end{equation}}
\newcommand{\ba}{\begin{array}}
\newcommand{\ea}{\end{array}}
\newcommand{\beqa}{\begin{eqnarray}}
\newcommand{\eeqa}{\end{eqnarray}}
\newcommand{\beqs}{\begin{subequations}}
\newcommand{\eeqs}{\end{subequations}}

\def\({\left(}
\def\){\right)}

\def\leqq{\leqslant}
\def\geqq{\geqslant}
\def\End{\end{document}}

\begin{document}

\title{Inelastic Dark Matter and High-Energy Recoil Signatures in LZ}
\author{{\sc Zi-Tong Fan}}
\email{fanzt2006@sjtu.edu.cn}
\affiliation{State Key Laboratory of Dark Matter Physics, \\
Tsung-Dao~Lee Institute $\&$ School of Physics and Astronomy, \\
Shanghai Jiao Tong University, Shanghai, China}

\author{{ \sc Hong-Jian He}}
\email{hjhe@sjtu.edu.cn}
\affiliation{State Key Laboratory of Dark Matter Physics, \\
Tsung-Dao~Lee Institute $\&$ School of Physics and Astronomy, \\
Shanghai Jiao Tong University, Shanghai, China}
\affiliation{Department of Physics, Tsinghua University, Beijing, China; \\
Center for High Energy Physics, Peking University, Beijing, China}

\author{{\sc Yu-Chen Wang}}
\email{wang-yc15@tsinghua.org.cn}
\affiliation{State Key Laboratory of Dark Matter Physics, \\
Tsung-Dao~Lee Institute $\&$ School of Physics and Astronomy, \\
Shanghai Jiao Tong University, Shanghai, China}

\author{{\sc Yue Zhao}}
\email{zhaoyue.hep@gmail.com}
\affiliation{Department of Physics and Jockey Club Institute for Advanced Study,\\
The Hong Kong University of Science and Technology, Hong Kong, China}

\begin{abstract}
We study the inelastic dark matter (iDM) that consists of two-component dark matter particles
$(\Xa,\Xb)$ and serves as a minimal extension beyond the commonly used one-component DM models.\ 
A distinctive feature of such iDM scattering with the target nuclei is to favorably produce 
signals at high nuclear recoil (NR) energy region in both exothermic and endothermic processes.\ 
In particular, for the exothermic dark matter scenario, the signal does not rely on the tail of the Boltzmann velocity distribution. As a result, the required DM-nucleus scattering cross section is significantly reduced, which in turn helps to substantially relax the stringent constraints imposed by IceCube neutrino searches.\ 
Using the inelastic DM-nucleus scattering, we naturally explain 
the newly reported event excess at high recoil energy with the LUX-ZEPLIN (LZ) experiment.
\\[2mm]  
Chin. Phys. Lett. 43 (2026) 110201 (Express Lett.) {[{\hs}arXiv:2609.10491{\hs}]}.  
\end{abstract}
\maketitle

\section{\hspace*{-3mm}Introduction}
\label{sec:introduction}
\label{sec:1}

Identifying the particle nature of dark matter (DM) poses a central challenge
in particle physics and cosmology.\ Direct DM detection provides a fundamental  
means to overcome this challenge by searching for the energy deposited
when Galactic DM scatters on ordinary matter.\ A confirmed signal would
establish a nongravitational interaction of DM and offer information
about its mass and couplings.\ This prospect has motivated extensive
experimental efforts so far, including LZ\,\cite{LZ2025}, XENON1T\,\cite{XENON1T2024}, 
and PandaX\,\cite{PandaX2025}, which have placed
stringent constraints on DM-nucleon scattering.

For conventional weakly interacting massive particles (WIMPs), these
searches primarily target low-energy nuclear recoils.\ In the standard
picture of momentum-independent, elastic scattering,
the recoil energy is supplied entirely by the kinetic energy of
nonrelativistic halo particles.\ Larger recoils require faster incident
particles, while the nuclear form factor further suppresses scattering
at large momentum transfer.\ The resulting spectrum is thus 
concentrated at low energies and falls rapidly toward higher energies.

The recent LZ observation\,\cite{LZ2026} presents an unusual departure from this
expectation.\ In a search extending the nuclear-recoil energy window to
approximately $270\keV$, the collaboration reported one event consistent
with a nuclear recoil of
$E_R^{}\!=\! 248\!\pm\!23\,\mathrm{(stat)}\!\pm\!23\,\mathrm{(sys)}\keV$ in a region
with a low expected background\,\cite{LZ2026}.\ The analysis finds a 
local significance of $3.4\hs\sigma$ across the tested models and a global
significance of $2.6\hs\sigma$.\ The challenge for a conventional WIMP
interpretation is not simply the large recoil energy, but its occurrence 
without a corresponding low-energy excess.\ Increasing the elastic
scattering cross section to account for the high-energy event would simultaneously
enhance the more abundant low-energy recoils.\ Hence, a DM explanation 
calls for a mechanism that can properly change the shape of the recoil spectrum,
instead of merely its normalization.

Inelastic scattering\,\cite{TuckerSmith2001} offers such a mechanism
and has motivated several interpretations of the LZ event\,\cite{
Su2026,Fan2026,Freese2026,Lou:2026idn,Nomura2026,DiMauro2026,Visinelli:2026kgt,Yamashita2026,Smirnov2026,DuWang2026,WuZhangZhu2026,Rodd2026,McCabe2026,DentNewstead2026,deLima2026,BaerBarger2026,
Lee:2026wof,Wang:2026ytg,Yang:2026wpb,Kotlarski:2026pep,DiMauro:2026dqp,Das:2026uyy,Alhazmi:2026efz,Okada:2026eol,Ahmed:2026qjg,Du:2026lpa,Bandyopadhyay:2026gjw,Kannike:2026qyl,Borah:2026zwf,Bose:2026ndd,Bisal:2026khf,
Cheung2026,Yuan2026,Zhu2026,Asadi2026,Lee2026,Lee2026b,Langhoff:2026ujr,Nguyen:2026lui,Lian:2026hpm,Lian:2026bze}, 
whereas other types of interpretations have been discussed\,\cite{Jeesun:2026vzo,Gu:2026vto,Liang:2026coz,Elahi2026,Aghaie2026,Unwin:2026rdp}.\ 
An inelastic process can be either endothermic or exothermic.\ 
In an endothermic process, an incident DM particle scatters into a heavier state.\ 
The excitation energy introduces a kinematic threshold that suppresses low-energy recoils and
can shift the signal toward the observed energy.\ 
However, many well-motivated realizations of such a model, like Higgsino in supersymmetry, 
are subject to stringent constraints from searches for high-energy
neutrinos from the Sun\,\cite{PospelovRamani2026}.\ 
The large cross section in the endothermic scenario predicts 
large neutrino flux from DM annihilation inside the Sun.
Thus, the absence of the corresponding IceCube signal\,\cite{IceCube2025} excludes 
the thermal Higgsino interpretation of the LZ event.\

One the other hand, the exothermic inelastic scattering\,\cite{Graham2010,He2021EFT,He2021Vector,He2024,Wang2025} also presents high-energy recoils by injecting the energy from the mass-splitting into the final-state kinetic energy and produces signals compatible with the LZ event\,
\cite{DentNewstead2026,deLima2026,BaerBarger2026}. 
For suitable masses and splittings, this
process generates a characteristic recoil scale at high energy and 
kinematically suppresses the recoils near the zero energy.\ Hence, it can 
produce the spectral pattern indicated by the LZ anomaly without relying
solely on the high-speed tail of the halo. Particularly, for the same DM mass, mass-splitting (with opposite signs) and cross section, 
the exothermic process always induces stronger signal than the endothermic process
due to kinematic enhancement. 
This requires lower interaction strength to explain the LZ data, 
and so it is easier to evade other constraints. 
Exothermic kinematics, together with a
sufficiently suppressed solar-neutrino yield, thus offers a route to
addressing both the recoil-spectrum problem and the indirect-detection tension.\ 

In this work, we investigate both exothermic and endothermic inelastic DM as an attractive 
interpretation to the observed event excess in the LZ experiment.\ 
In Sec.\,\ref{sec:2} we study the inelastic scattering mechanism in a minimal two-state setup
with a short-range, spin-independent interaction,  
and calculate the expected nuclear recoil event rate in LZ. 
We examine its compatibility with the
LZ candidate in Sec.\,\ref{sec:3}.\ 
We also discuss the
requirements for a surviving excited-state population and the
conditions under which solar-neutrino constraints can be avoided in Sec.\,\ref{sec:4}.
Finally, we conclude in Sec.\hs\ref{sec:5}.

\section{\hspace*{-3mm}Theoretical Setup for iDM}
\label{sec:2}
\label{sec:model}

In this section, we first set up the minimal iDM 
and discuss its kinematic properties.\ 
Then, we present the effective iDM-nucleon interaction and calculate
the inelastic nuclear recoil rates. 

\vspace{-3mm}
\subsection{\hspace*{-3mm}Minimal Inelastic Dark Matter}
\label{sec.2.1}
\label{sec:kinematics}
\vspace{-2mm}

We consider two nearly degenerate Dirac fermions, $\chi_1^{}$ and $\chi_2^{}\hs$,
with a common mass scale $m_\chi^{}\!\simeq\! m_{\chi_1}\!\simeq\! m_{\chi_2}$.\ 
For the transition $\chi_1^{}\!+N\!\to\hsm\chi_2^{}\hsm +N$, we define 
the DM mass splitting,
\\[-7mm]
\begin{equation}
\Delta m =  m_{\chi_1}^{}\!\!-m_{\chi_2}^{}\hs.
\label{eq:splitting}
\end{equation}
Thus, the case of $\Delta m\!>\!0$ denotes the exothermic scattering, whereas 
the case of $\Delta m\!<\!0$
corresponds to the endothermic scattering.\ 
In either case,  $\chi_1^{}$
always labels the incident state. 

\hs 

We denote the target nuclear mass by $m_N^{}$ and
$\mu_N^{}\!=m_\chi m_N/(m_\chi\!+\!m_N)$ is the reduced mass.\ 
The energy and momentum
conservations in the center-of-mass frame give the following conditions:
\begin{equation}
\hspace*{-4mm}
\frac12\mu_N v^2\!+\!\Delta m=\frac12\mu_N v'^2,
~~~~
\bm q=\mu_N(\bm v\!-\!\bm v')\hs,
\label{eq:energy_conservation}
\end{equation}
where $\bm v$ and $\bm v'$ are the relative velocities of 
the initial and final states respectively.\ 
The nuclear recoil energy is $E_R^{}\!=\!q^2/(2m_N)$.\ 
Then, the minimal relative velocity kinematically allowed
is given by
\begin{equation}
v_{\min}(E_R)=\frac{1}{\sqrt{2m_NE_R}}
\left|\frac{m_N E_R}{\mu_N}-\Delta m\right|.
\label{eq:vmin}
\end{equation}

For exothermic scattering, $v_{\min}^{}$ vanishes when the nuclear recoil energy
$E_R^{}\!=\!\mu_N^{}\Delta m/m_N^{}$, 
where an enhancement of signal is expected.\  
Also, for any nonzero $\Delta m$,
$v_{\min}^{}\hsm\!\sim\!\Delta m/\!\sqrt{2m_N^{}E_R^{}}$ diverges as 
$E_R^{}\hsm\!\to\! 0\hs$.\ 
This suppresses low NR events and makes the exothermic process relevant to the LZ event. 

\vs 

For endothermic scattering, 
the minimum of $v_{\min}(E_R)$ occurs at 
$E_R^{}\!=\!\mu_N|\Delta m|/m_N$ 
and this minimum equals $\sqrt{2|\Delta m|/\mu_N^{}\,}\hs$.\ 
The energy conservation \eqref{eq:energy_conservation}
imposes an upper bound on the DM mass-splitting in the case,
\begin{equation}
|\Delta m|<\frac12\mu_Nv_{\max}^2 \,. 
\label{eq:endo_threshold}
\end{equation}

At recoil energy around $E_R^{}\!=\!\mu_N^{}\Delta m/m_N^{}$, 
we always have the following inequalities for the minimal relative incident velocities
in the exothermic, elastic and endothermic scattering cases,
\begin{equation}\label{eq:vmin-comparison}
v_{\min}^{\rm exo}<v_{\min}^{\rm elastic}<v_{\min}^{\rm endo}.     
\end{equation}
Hence, for producing the same event rate, 
the exothermic scattering requires the weakest interaction strength,
whereas the endothermic scattering requires the strongest.

\begin{figure*}[t]
\centering
\includegraphics[height=6cm]{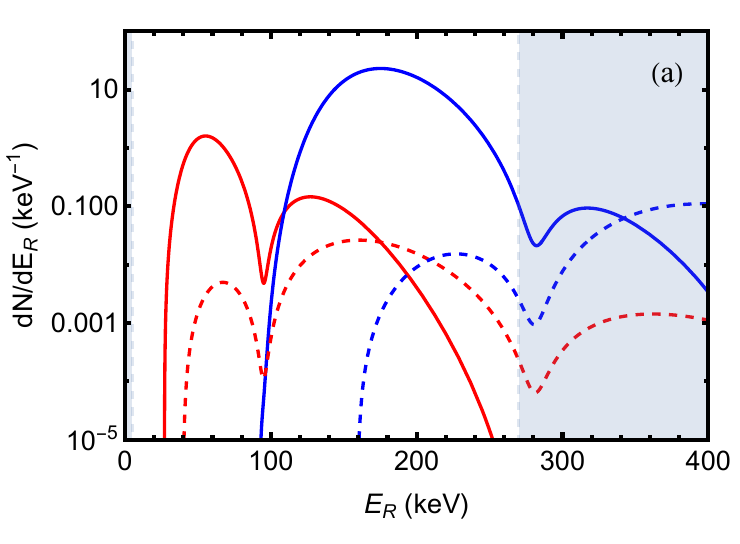}
\includegraphics[height=6cm]{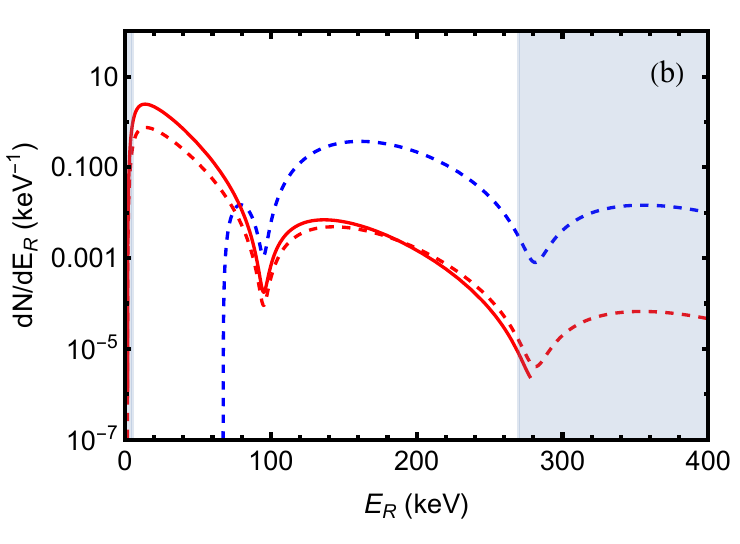}
\vspace*{-5mm}
\caption{%
Plot\,(a): Spectra of true event number for the exothermic iDM.\  
The solid (dashed) curves represent DM mass $m_\chi\!=\!50\hs$GeV\,(500{\hs}GeV).\ 
The red curves are for $(\Delta m, \sigma_p)\!=\!(300{\rm keV},10^{-45}{\rm cm^2})$, 
    whereas the blue curves are for $(\Delta m, \sigma_p)=(700{\rm keV},10^{-43}{\rm cm^2})$. 
Plot\,(b): Spectra of true event number for the endothermic iDM.\  
The solid (dashed) curves represent DM mass $m_\chi\!=\!200\hs$GeV (900{\hs}GeV), 
The red curves are for $(\Delta m, \sigma_p)\!=\!(-50\hs{\rm keV},10^{-45}{\rm cm^2})$, 
whereas the blue curves are for 
$(\Delta m, \sigma_p)\!=\!(-250{\rm keV},10^{-41}{\rm cm^2})$.\  
The gray shaded regions at $E_R^{}\hsm\!<\!5\hs$keV 
and $E_R^{}\hsm\!>\!270\hs$keV 
indicate where the detection efficiency is below 50\%{\hs}.\ 
Note that there is no blue solid curve in plot\,(b), 
because it is kinematically forbidden.}
\label{fig:spectra}
\label{fig:1}
\end{figure*}

\subsection{\hspace*{-3mm}Interaction and Nuclear Recoil Spectrum}
\label{sec:2.2}
\label{sec:spectrum}

We formulate the spin-independent iDM-nucleon interaction 
by an off-diagonal vector-type effective Lagrangian,
\begin{equation}
\mathcal L_{\rm eff}=
\frac{1}{\Lambda^2}\!\sum_{a=p,n}\!\!c_a^{}
(\bar\chi_2\gamma^\mu\chi_1)(\bar a\gamma_\mu a)
+\mathrm{h.c.}
\label{eq:effective_operator}
\end{equation}
%
$\chi_1$ and $\chi_2$ can also represent four-component Majorana fermions, 
in which case the coefficients $c_a$ are purely imaginary.\ 
We note that the diagonal terms can be absent and 
thus tree-level elastic scattering vanishes, 
as shown by the UV-completion theories in the literature\,\cite{He2021Vector,He2024}.\ 

\vs 

We define the reference iDM-proton scattering cross section
and the corresponding nuclear normalization as follows: 
\\[-6mm]
\beqs 
\begin{align}
\sigma_p^{} &= \frac{~\mu_p^2|c_p|^2~}{\pi\Lambda^4} \hs,
\label{eq:sigma_proton}
\\
\sigma_N^0 & = \sigma_p^{}\frac{\,\mu_N^2\,}{\mu_p^2}\!\!
\left|Z\!+\!(A\!-\!Z)\frac{c_n}{c_p}\right|^2 ,
\label{eq:sigma_nucleus}
\end{align}
\eeqs 
where $A$ and $Z$ denote the atomic mass and charge respectively.\ 
We consider the conventional isospin symmetry between proton and neutron, 
which gives $c_n^{}\!=\!c_p^{}$ and the familiar coherent factor $A^2$
in Eq.\eqref{eq:sigma_nucleus}.\ 
The differential recoil cross section is given by
\begin{equation}
\frac{\dd\sigma_N}{\dd E_R}
=\frac{m_N^{}\sigma_N^0}{~2\hs\mu_N^2v^2~}F^2(q)\,, 
\label{eq:differential_cross_section}
\end{equation}
where the nuclear response is modeled with the Helm form factor, 
\begin{equation}
\begin{gathered}
\hspace*{-5.4mm}
F^2(q)=
\left[\!\frac{\,3\hs j_1^{}(q\hs r_N^{})\,}
{q\hs r_N^{}}\!\right]^{\!2}
\!e^{-(qs)^2},
\\[1mm]
r_N^{}=1.14A^{1/3}\,\mathrm{fm}\hs, \quad 
s=0.9\,\mathrm{fm}\hs.
\end{gathered}
\label{eq:helm}
\end{equation}
The above formulas hold for the non-relativistic limit
and the same formulas apply to the case of scalar iDM. 

\vs 

The typical momentum transfer is $q_0^{}\!\simeq\hsm 0.25\GeV$
around the recoil energy $248\keV$.\ 
For a mediator with a heavy mass $m_V^2\!\gg\! q_0^2\hs$,
the propagator induces a factor of $m_V^{-4}$, 
relatively suppressing low energy recoils as compared to the light-mediator case  
where the propagator contribution scales as 
$q^{-4}\!\propto\hsm E_R^{-2}$. 

\vs 

Let $\rho_{\rm in}^{}$ be the density of the incident DM particles.\ 
The differential rate per unit detector mass is given by 
\begin{equation}
\frac{\dd R}{\dd E_R}
=\frac{\rho_{\rm in}}{m_\chi}
\frac{\sigma_N^0}{2\mu_N^2}F^2(q)
\hsx\eta\hsm\(v_{\min}(E_R)\)\hs ,
\label{eq:rate}
\end{equation}
with the function 
\begin{equation}
\eta(v_{\min})=\int_{v\geqq v_{\min}}^{}\hspace*{-5mm}\dd^3v\hs 
\frac{\,f_{\rm gal}(\bm v\!+\!\bm v_e^{})\,}{v}\,.
\label{eq:eta}
\end{equation}
The velocity distribution $f_{\rm gal}(\bm v)$ is a Maxwell-Boltzmann distribution truncated
at the escape velocity $v_{\rm esc}^{}$ and normalized to unity.\ 
For illustration, we adopt the Standard Halo Model (SHM) with 
inputs $\rho_{\rm DM}^{}\!=\!0.3\GeV\,\mathrm{cm}^{-3}$,
$v_0\!=\!v_e\!=\!0.00073c\hs$, and $v_{\rm esc}=0.00181c$, 
neglecting annual modulation\,\cite{McCabe:2010zh}.\  
Thus, $v_{\max}^{}\!=\!v_{\rm esc}^{}\!+\!v_e\!=\!0.00254c\hs$.\ 
Without loss of generality, we choose 
$\rho_{\rm in}^{}\!=\!\rho_{\rm DM}^{}$ henceforth. 

\vs 

We obtain the true event spectra by 
multiplying Eq.\eqref{eq:rate} with the exposure of the LZ experiment, 
$\mathcal{E} \!=\! 2.84$ tonne-years.\  
In Fig.\,\ref{fig:spectra}(a) we show the exothermic case where $\Delta m\!>\!0\,$.\  
It is clear that the larger $m_\chi^{}$ and larger $\Delta m$ 
give higher $E_R$ thresholds, 
and are more compatible with the event excess at $E_R\!\sim\! 248\hs$keV.\ 
However, the larger $(m_\chi, \Delta m)$ values also predict 
much more events in the higher NR region, 
and will be constrained by the higher-energy sideband (HE\,SB).\footnote{%
Higher-energy sideband information is particularly important when the
spectrum extends beyond the main acceptance.\ 
The potential importance of this test has also been emphasized in Refs.\,\cite{Rodd2026,DentNewstead2026}.\ 
Without further information of the sideband, 
in this work we do not establish a sideband exclusion 
or an upper bound on the DM mass.}\ 
The endothermic iDM ($\Delta m\hsm\!<\hsm\!0$) in Fig.\,\ref{fig:1}(b) 
also shows that larger $m_\chi$ and larger $|\Delta m|$ give higher $E_R$ thresholds.\  
We note that there is no blue solid curve in plot\,(b).\  
This is because it violates the condition \eqref{eq:endo_threshold}.

\section{\hspace*{-3mm}Analyzing iDM Signatures for LZ}
\label{sec:3}

In this section, we study the iDM signal in LZ detectors
and perform a simplified profile likelihood analysis to set limit for 
$m_\chi$, $\Delta m$ and $\sigma_p$. 

\vspace*{-1.5mm}
\subsection{\hspace*{-3mm}Detector Response and Statistical Methods}
\label{sec:3.1}
\label{sec:likelihood}
\vspace*{-1.5mm}

We use the efficiency $\epsilon(E_R)$
digitized from Fig.\,S2 of Ref.\,\cite{LZ2026} and the exposure
$\mathcal{E}\!=\!2.84$ tonne-years to calculate the event numbers: 
\begin{equation}
\hspace*{-3mm}
\frac{\dd N}{\dd E_d}
=\hs\mathcal{E}\!\hsm\int_0^\infty\!\!\!\dd E_R^{}\,
G(E_d^{};E_R^{},\sigma_E^{}(E_R^{}))\hs\epsilon(E_R^{})
\frac{\dd R}{\dd E_R} \hs,
\label{eq:response}
\end{equation}
where $E_d^{}$ is the detected energy, 
$G$ a normalized Gaussian kernel and $\sigma_E^{}$ the energy resolution.\  
The uncertainty can be estimated as
$\sigma_E^{}\!=\!\sqrt{\sigma_{\rm stat}^2
\!+\hsm\sigma_{\rm sys}^2\,}\,$,
with 
\begin{equation}
\sigma_{\rm stat}(E_R) 
= 1.46\sqrt{E_R\,},
~~~
\sigma_{\rm sys}(E_R) 
= 0.093E_R\,,
\label{eq:resolution}
\end{equation}
where all quantities are in the unit of keV.

\begin{figure*}[t]
\centering
\hspace*{-3mm}
\includegraphics[width=9cm,height=6.5cm]{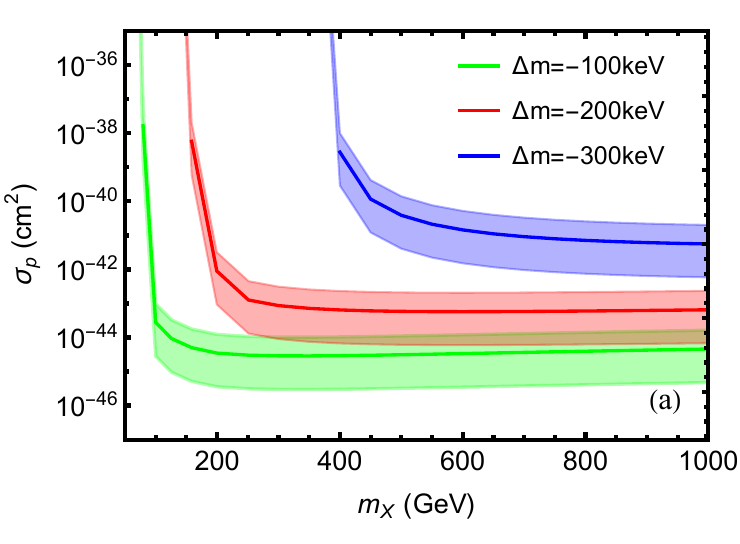}
\hspace*{-2mm}
\includegraphics[width=9cm,height=6.5cm]{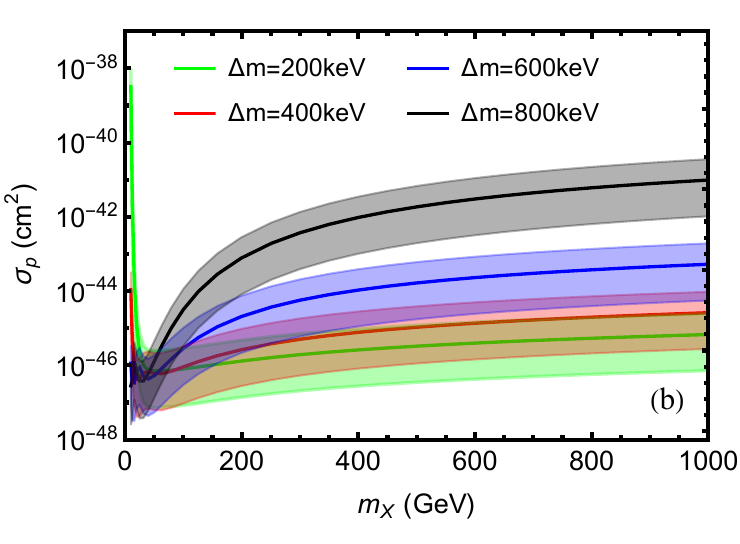}
\vspace*{-4mm}
\caption{%
Best fits (solid curves) and 90\%\,C.L.\ intervals (shaded bands) 
on the DM-proton scattering cross section.\  
Plot\,(a) Limits on $\sigma_p$ versus $m_\chi$ for endothermic iDM, 
with benchmark values of $\Delta m$ as stated in the plot legend.\  
Plot\,(b) Limits on $\sigma_p$ versus $m_\chi$ for exothermic iDM, 
with benchmark values of $\Delta m$ as stated in the plot legend.}
\label{fig:2}
\label{fig:results-sigma-mx}
\end{figure*}

\vs 

We use the three intervals $[70,215.5]$, $[215.5,280.5]$, and $[280.5,305]\keV$ 
for $E_d$ binning and assign the event numbers $(n_1,n_2,n_3)\!=\!(0,1,0)$.\  
For the lower bound of $E_d^{}\,$, we adopt from Fig.\,2 of Ref.\,\cite{LZ2026}, $E_d^{\rm min} \!\!=\!\! 70\,$keV, 
above which the electron-recoil/nuclear-recoil discrimination 
becomes effective.\ 
The upper bound $E_d^{\rm max}\!=\!305$\,keV is simply 
derived by $270\,{\rm keV}\!+\!\sigma_E^{}(270\,{\rm keV})$, 
where the value 270\,keV is the nuclear recoil energy 
at which the efficiency falls below {50\%\hs}.\ 
The total background count is only $0.0106\pm 0.0008$\,\cite{LZ2026},  
and thus is neglected for simplicity in the following.\  

\vs 

For fixed $(m_\chi,\Delta m)$ values, we write the expected signal counts as
$\mu\hs s_i^{}\hs$, where $\mu$ is the dimensionless parameter for the signal strength
and $s_i$ is the predicted event number in the $i$-th bin (for a benchmark $\sigma_N$).\ 
The background-free likelihood is given by
\begin{equation}
\hspace*{-4mm}
\mathcal L(\mu;m_\chi,\Delta m)=\mu s_2e^{-\mu s_{\rm tot}},
~~~ s_{\rm tot}=\sum_{i=1}^3s_i^{}\,.
\label{eq:likelihood}
\end{equation}
This likelihood function can be maximized analytically with respect to $\mu\hs$.\  
It gives the best fit value $\hat\mu\hs$, 
and the corresponding profiled likelihood 
$\mathcal L_{\rm{prof}}$ for $(m_\chi,\Delta m)$: 
\begin{equation}
\hspace*{-4mm}
\hat\mu(m_\chi,\Delta m)=\frac{1}{\,s_{\rm{tot}}^{}\,},
~~~
\mathcal L_{\rm{prof}}^{}(m_\chi^{},\Delta m)
=\frac{s_2^{}}{\,s_{\rm tot}e\,} \hs .
\label{eq:profile}
\end{equation}
Thus, the fitted total signal count is always one, 
and the fit quality affected by the spectral shape 
is represented solely by the fraction of the accepted signal in the event-containing interval.

\vs 

For the shape comparison, we define the test statistic
\begin{equation}
\chi^2(m_\chi^{},\Delta m) = -2\ln(\mathcal L_{\rm prof}^{})\,,
\label{eq:relative_likelihood}
\end{equation}
to perform the likelihood-based test for limit setting.

\subsection{\hspace*{-3mm}Analyzing iDM Signatures at High Recoil Energy}
\label{sec:3.2}
\label{sec:parameters}

\begin{figure}[b]
\centering
\hspace*{-3mm}
\includegraphics[width=8.9cm,height=6.3cm]{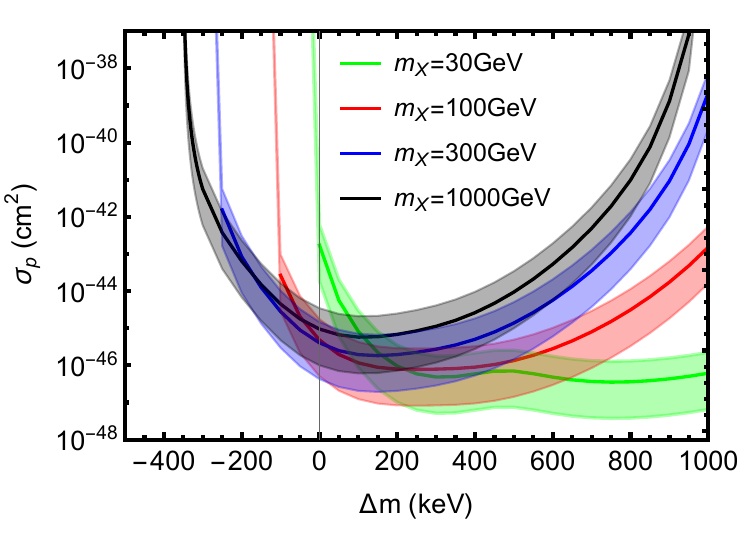}
\vspace*{-8mm}
\caption{%
Best fits (solid curves) and 90\%\,C.L.\ intervals (shaded bands) on the DM-proton scattering 
cross section versus $\Delta m$ for both endothermic and exothermic iDM.\  
The benchmark values of $m_\chi$ as depicted in the plot legend. 
}
\label{fig:results-sigma-dm}
\label{fig:3}
\end{figure}

We scan over the parameter space of 
$10\GeV\!\leqq\! m_\chi\!\leqq\!1000\GeV$ 
and $-500\keV\!\hsm\leqq\!\hsm\Delta m
\hsm\!\leqq\hsm\!1000\keV$, 
excluding kinematically inaccessible regions 
where the condition \eqref{eq:endo_threshold} is not satisfied. 
The upper boundary for $m_\chi$ is to avoid excesses 
in the higher-energy sideband, 
and the upper boundary for the mass-splitting is simply $\Delta m<2m_e$. 
The lower boundaries for $m_\chi$ and $\Delta m$ are for kinematic reasons. 

\vs 

Fig.\,\ref{fig:results-sigma-mx} shows the best fit results and 90\% C.L. intervals 
for the DM-proton scattering cross section $\sigma_p$ versus DM mass $m_\chi$. 
(For simplicity, in converting the DM-nucleus scattering cross section $\sigma_N$ to $\sigma_p$, 
we adopt the conventional assumption of a isospin symmetry between proton and neutron.)
In plot (a) for endothermic scattering, 
we see that each value of $|\Delta m|$ sets a cutoff on $m_\chi$ from below. 
This can be derived from the kinematic condition in Eq.\eqref{eq:endo_threshold}, 
which implies
\begin{equation}
\hspace{-4mm}
\mu_N^{}v_{\max}^2\!>\hsm 2|\Delta m|\hs, 
~~~ m_\chi\!>\!
\frac{2m_N|\Delta m|}{\,m_N^{}v_{\max}^2\!-\!2|\Delta m|\,} \,. 
\label{eq:mass_threshold}
\end{equation}
The scan also contains a broad exothermic region with favorable spectral
likelihood, as shown in plot-(b).\ 
In both plots, the reduced masses and spectral shape approach constants 
for $m_\chi^{}\!\gg\! m_N^{}\hs$, 
whereas the incident number density remains proportional to $m_\chi^{-1}$.\  
Comparing the results between plot (a) and (b), 
we see that the exothermic case in general requires much smaller cross section to fit the event.\  
This is explained by Eq.~\eqref{eq:vmin-comparison}. 

\begin{figure}[t]
\centering
\hspace*{-3mm}
\includegraphics[width=9.1cm,height=6.7cm]{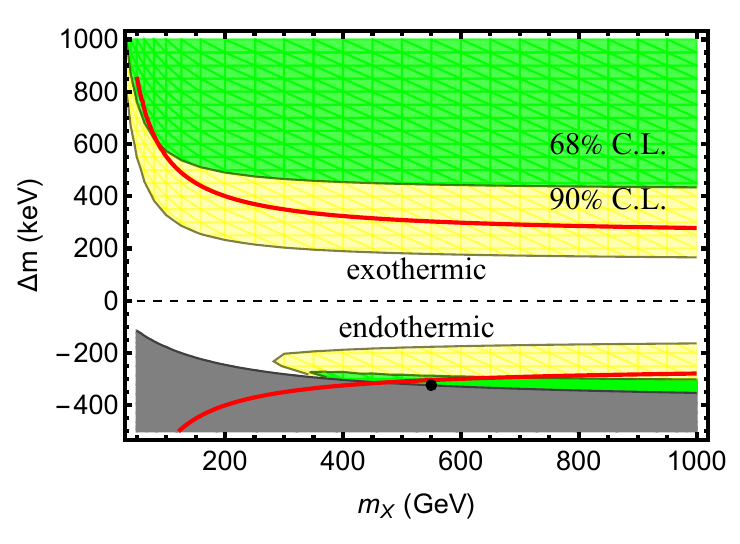}
\vspace*{-8mm}
\caption{%
Best fit value (black dot), 68\% \,C.L.\ limits (green regions) and 90\%\,C.L.\ 
limits (yellow regions) for the DM mass $m_\chi^{}$ and mass-splitting $\Delta m\hs$.\  
The gray shaded region is kinematically forbidden.\  
The red curves represent the case where a peak at 
$E_R^{}\!=\!248\,$keV is predicted.\ 
}
\label{fig:results-contour}
\label{fig:4}
\end{figure}

\vs 

We plot the best fit results and 90\% C.L.\ intervals 
for the DM-nucleon scattering cross section $\sigma_p^{}$ 
versus the mass-splitting $\Delta m$ in Fig.\,\ref{fig:results-sigma-dm}. 
The left region with $\Delta m<0$ stands for the endothermic scattering, 
whereas the right region with $\Delta m\!>\!0$ is for the exothermic scattering. 
For each value of $m_\chi$, there is a lower limit on $\Delta m$, 
which is exactly determined by Eq.\eqref{eq:endo_threshold}. 

\vs 

In Fig.\,\ref{fig:results-contour}, 
we show the 68\%\,C.L. and 90\%\,C.L.\ contours on the 
$m_\chi\!-\!\Delta m$ plane.\  
The red curves represent 
$|\Delta m|\!=\!(m_N^{}/\mu_N^{})\hsm\times\hsm 248\,{\rm keV}$, 
in which case a peak at $E_R^{}\!=\hsm 248$\,keV is predicted.\  
The gray shaded area is kinematically forbidden 
by violating the condition \eqref{eq:endo_threshold}.\  
The scan places its largest likelihood near 
$(m_\chi,\Delta m)\!\simeq\!(550\GeV,-323\keV)$.\ 
The best fit point is understandable as a near-boundary solution rather than
a well-determined iDM benchmark.\  
At the boundary we have 
$|\Delta m|\!=\!\mu_Nv_{\max}^2/2$. 
Setting $E_R\!=\!\mu_N|\Delta m|/m_N\!=\!248\,$keV gives 
$m_\chi^{}\!\sim\!466\,$GeV and 
$|\Delta m|\hsm\sim\hsm 313\,$keV, 
which explains the best fit values.\  
This boundary point gives a narrow spectrum whose shape fits the excess very well, 
but its rate tends to zero.\ 
To fit the observed event using the boundary point requires 
a divergent cross section, thus this best-fit point 
is not physical.\  
Similarly for the exothermic case, as discussed in Sec.\,\ref{sec:2}, 
the larger $(m_\chi, \Delta m)$ values produce much more events in the higher NR energy region ($\geqq\,$300\,keV), 
and the observed event at 248\,keV corresponds to the low-energy tail of the full spectrum.\ 
Hence, a much larger total event number (and thus a much larger cross section) is expected to explain this event.\ 
Such large values of $(m_\chi, \Delta m)$ should be constrained 
once the HE\,SB data is released, 
and we expect that the red curve would become more favored 
by data.\ 
Despite of these unphysical regions, 
the most parameter space inside the 90\%\,C.L.\ and 90\%\,C.L.\ contours 
belongs to the exothermic case.\ 
We further note that at 68\%\,C.L.\ 
the endothermic case favors 
$165\,{\rm keV}\!<\!|\Delta m|\!<\!350\,{\rm keV}$, 
whereas the exothermic case favors 
$165\,{\rm keV}\!<\!\Delta m\!<\!1000\,{\rm keV}$.

\vspace*{-4mm}
\section{\hspace*{-3mm}Cosmological and Astrophysical Constraints}
\label{sec:4}
\vspace*{-2mm}

In this section, we present the cosmological and astrophysical constraints
on the iDM scenario.\ 
We first discuss the heavier-state abundance and its longevity.\ 
Then, we estimate the solar capture of the exothermic and endothermic iDM 
as well as study the total annihilation rates.\
We show that the iDM annihilation-induced solar neutrino signals are sufficiently suppressed and well below the IceCube neutrino bound.

\vspace*{-2mm}
\subsection{\hspace*{-3mm}Excited-State Abundance and Longevity}
\label{sec:4.1}
\label{sec:population}
\vspace*{-1.5mm}

Conventionally, it is assumed that DM consists of only the lightest particle in the dark sector. 
However in the inelastic case, since there are two nearly degenerate states, 
the constituent of DM may differ and need to be revisited. 
From a purely theoretical standpoint, 
an endothermic process always requires a sufficient up-scattering cross section 
and a small enough mass-splititng $\Delta m < m_\chi v^2 \sim 10^{-6}m_\chi$. 
Thus the two states must have been 
in thermal and chemical equilibrium in the early Universe 
and then freeze out with nearly equal abundances. 
If the heavier state is sufficiently long-lived, 
it should constitute a significant fraction of the local dark matter today. 
Consequently, the exothermic process is kinematically favored 
and could be the dominant contribution.
Therefore, a proper treatment of the exothermic scattering is essential 
for a complete description of direct detection signals. 

\vs 

Throughout the analysis, 
we always study the endothermic and exothermic inelastic scatterings separately.\  
This is because as long as the heavier component is as abundant as the lighter one, 
the exothermic process dominates because it is enhanced kinematically, 
as discussed by Eq.\eqref{eq:vmin-comparison}.\ 
From Fig.\,\ref{fig:2}, we see that for $|\Delta m|=200$\,keV 
the endothermic case [the red curve in plot\,(a)] requires 
a cross section $(10^2\hsm\! -\hsm\! 10^6)$ times larger than the exothermic case 
[the green curve in plot\,(b)].\ 
Hence, to obtain comparable contributions from the endothermic and exothermic processes 
(having identical cross sections), it requires 
$\rho_\chi^{\rm heavy}\!/\rho_\chi^{\rm light}
\!=\! 10^{-2}\!-\!10^{-6}$.\ 
For simplicity, we do not consider such a scenario in this work.\ 
Ref.\,\cite{BaerBarger2026} has discussed the case of 
$\rho_\chi^{\rm heavy}\!/\hsm\rho_\chi^{\rm light} \!\sim\!10^{-3}$, 
and showed that the exothermic process is still dominant.\ 
Hence, as long as the light and heavy DM particles coexist with comparable abundance, 
the main effect always comes from the heavier iDM component and
we only need to take into account the exothermic scattering 
in direct detection experiments.\ 
This could be an additional reason to motivate 
the exothermic scenario.

\vs 

However, comparable populations do not follow from near-degenerate masses alone.\ 
Firstly, a long decay lifetime is necessary for a primordial excited population. 
For the dark photon-mediated iDM model,
the heavier state dominantly decays  
into the lighter state plus a neutrino-pair with the decay width\,\cite{He2024}:  
%
\begin{eqnarray}
\small
\hspace*{-8mm}
\Gamma_{\chi_1\rightarrow \chi_2\nu\bar\nu}^{} & \sim &
\left(4\!\times\!10^{34}\text{yrs}\right)^{\!-1}\!\!
	\(\!\!\frac{\Lambda}{\,100\hs\text{GeV}\,}\!\)^{\hsm\!\!-4}\!\!\hsm
	\(\!\frac{\Delta m}{\,10\hs\text{keV}\,}\!\)^{\!\!9}
\nonumber\\[1mm]
\hspace*{-8mm}
& = &  
\(\hsm 10^{25}\text{yrs}\)^{\!-1}\!\!
\left(\frac{\Lambda}{\,\rm{TeV}\,}\!\right)^{\hsm\!\!-4}
\!\!\hsm 
\(\hsm\!\frac{\Delta m}{\,300 \text{keV}\,}\!\hsm\)^{\!\!9} ,
\hspace*{4mm}
\end{eqnarray}
%
where $\Lambda$ is the cutoff scale derived from the fitting.\ 
We see that the lifetime of the heavier state is 
substantially far beyond the age of the present Universe.\  
Note that factors such as the fraction of the heavier state and the ratio between $c_p$ and $c_n$ will be taken into account for $\Lambda$; 
yet these factors do not change the conclusion.\  
Secondly, other processes such as $\chi_1^{}\!+\!\chi_1^{}\!\leftrightarrow\!\chi_2^{}\!+\!\chi_2^{}$ 
and $\chi_1\!+\hsm e\!\leftrightarrow\!\chi_2\!+\hsm e$ 
also deplete the abundance of the heavier state 
when the DM temperature $T_\chi$
falls below $|\Delta m|\,$.\ 
As long as these processes decouple at higher temperature $T_\chi\!\gg\!|\Delta m|\,$, 
such depletion effects are negligible\,\cite{He2021EFT,He2021Vector}.\ 
Since the thermal history is much more model-dependent than the lifetime, we would leave such detailed analyses to future work. 

\vspace*{-2mm}
\subsection{\hspace*{-3mm}Solar Capture and IceCube Constraints}
\label{sec:4.2}
\label{sec:solar}

As discussed in the Introduction, the solar-neutrino problem of the inelastic Higgsino is unusually severe~\cite{PospelovRamani2026}. The LZ recoil is generated close to the endothermic kinematic boundary, so the terrestrial rate samples the extreme Galactic velocity tail, whereas solar acceleration to $w\gtrsim1300~{\rm km\,s^{-1}}$ removes much of this kinematic suppression.
In addition, the Higgsino scattering strength is fixed by tree-level $Z$ exchange and its dominant annihilation mode is $W^+W^-$, so both capture and the hard-neutrino yield are predictive.

\vs 

\begin{figure}[t]
\centering
\hspace*{-3.5mm}
\includegraphics[width=9.2cm,height=6.8cm]{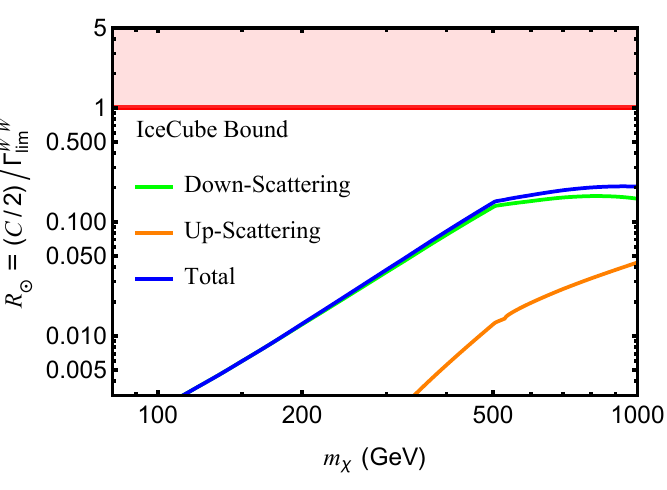}
\vspace*{-7mm}
\caption{%
The predicted exothermic iDM annihilation rate into 
$W^+W^-$ relative to the IceCube bound\,\cite{PospelovRamani2026,IceCube2025},
including heavier state down-scattering contribution (green line), lighter state up-scattering contribution (orange line) and total rate (blue line).}\ 
Across the DM mass range under consideration, 
the exothermic iDM model readily evades the IceCube neutrino bound at 90\%\,C.L (red line).\ 
The mass-splitting is derived from 
$\mu_{N}^{}\Delta m/m_{N}\!=\hsm 248\,$keV 
and the DM–nucleon cross section is fixed to the best fit of the LZ signal rate. 
\label{fig:solar}
\label{fig:5}
\end{figure}

Unlike the Higgsino scenario, solar-neutrino constraints are much weaker here. First, the exothermic DM scenario requires a significantly smaller scattering cross section with nucleus, since it does not rely on the velocity tail to reach the energy threshold. This drastically reduces the solar capture rate, such that even with dominant annihilation into electroweak bosons, IceCube imposes no constraint. Additionally, if annihilation does not predominantly produce electroweak bosons\,\cite{DiMauro:2026dqp}, the neutrino flux is further suppressed.\footnote{Ref.\,\cite{BaerBarger2026} considers the evasion solely due to the annihilation final state.\ But for exothermic iDM, the much smaller cross section lowers the solar capture rate enough to ensure a safe neutrino flux regardless of annihilation channels, 
as shown in Fig.\,\ref{fig:solar}.} Thus, the exothermic parameter space considered here is essentially unconstrained by current solar-neutrino searches.

\vs 
 
The kinematic distinction for exothermic capture follows directly from
energy conservation\,\cite{Blennow:2015hzp,Blennow:2018xwu}.\ 
Firstly, the endothermic process requires the kinematic threshold given by Eq.\eqref{eq:endo_threshold}.\  
For light nuclei such as hydrogen and helium (which make up about 98\% of the solar mass), 
the reduced mass is $\mu_N\!\approx\! m_N\!\approx\! O({\rm GeV})$, 
thus the up-scattering is forbidden even with the maximal solar escape velocity 
$v_{\rm{esc}}^{}\!\simeq\! 1380\,\rm{km}/\rm{s}
\hsm =\hsm 0.0046{\hs}c\,$.\ 
Secondly, a DM particle with asymptotic solar-frame speed $u$ has
speed $w$ at radius $r$, obeying the relation,
\begin{equation}
w^2=u^2+v_{{\rm esc},\odot}^2(r) \,.
\label{eq:solar_speed}
\end{equation}
Neglecting the thermal motion, an exothermic recoil leaves the DM with kinetic energy
$\tfrac12m_\chi w^2\!+\!\Delta m\!-\!E_R^{}$ at the leading order in $\Delta m/m_\chi\hs$.\ 
Requiring the outgoing state to remain gravitationally bounded thus gives
\begin{equation}
E_R^{}>\frac{1}{\,2\,}m_\chi^{} u^2\!+\hsm\Delta m\,.
\label{eq:exo_capture}
\end{equation}
Thus, although there is no kinematic threshold for the exothermic scattering, 
the extra energy from mass-splitting makes it harder to capture the iDM.\ 
By contrast, endothermic up-scattering is subject to a kinematic threshold, 
but the excitation itself removes kinetic energy and gives
$E_R\!>\!\tfrac{1}{2} m_\chi^{} u^2\!-\!\Delta m\hs$.\ 
Inside the Sun, the large local velocity $w$ can partially alleviate the endothermic excitation threshold.\ Hence, the extra released energy in the exothermic case  
tends to suppress its capture rate.

\vs 

We intentionally overestimate the solar neutrino signal 
as follows.\  
We adopt for all DM a low mono-speed 
$u_0\!\!=\!\!200\,{\rm km\,s^{-1}}$ which favors capture, 
and assume a one-zone core model for the solar composition, 
with the core potential 
$v_{{\rm esc},\odot}\!\!=\!\!1380\,{\rm km\,s^{-1}}$.\   
Using the formulas  
$E_R^{\pm}(w)\!\!=\!\!(\mu_N^2/2m_N^{})
(w\!\pm\hsm\!\sqrt{w^2\hsm+\hsm 2\Delta m/\mu_N^{}})^2$ and 
$w_0^2\!=\!u_0^2\!+\!v_{{\rm esc},\odot}^2$, 
we estimate the excited-state capture rate:
\begin{equation}
C_H\simeq
\frac{\rho_{\rm DM}}{m_\chi}\frac{w_0^2}{u_0} f_H
\sigma_p
\!\sum_i\! N_i
\frac{\,m_N^{}A_i^2\,}{\,2\mu_p^2w_0^2~}\!\!
\int_{E_R^{\rm cap}}^{E_R^+}\!\!\dd E_R\,F_i^2(q)\hs ,
\label{eq:capturerate}
\end{equation}
where $f_H^{}$ is the excited-state fraction, $i$ denotes the species of the target nucleus, 
$A_i$ the nuclear mass number, $N_i$ the number of target nuclei in the Sun, 
$F_i(q)$ is the Helm form factor, 
and we have $E_R^{\rm cap}\!=\!
\max\!\left[E_R^-,\,\Delta m\!+\!\frac12m_\chi u_0^2\right]$.\ 
We note that the cross section $\sigma_p^{}$ is already strongly constrained 
by the LZ data, as discussed in Sec.\,\ref{sec:3}. 
Given the inevitable presence of ground-state DM, we also estimate its solar capture rate by replacing $\Delta m$ with $-\Delta m$ and the excited-state fraction $f_H$ with the ground-state fraction $f_L$ in Eq.(\ref{eq:capturerate}).\ Following the discussion in Sec.\,\ref{sec:population}, we assume that the ground and excited states 
are equally abundant, $f_L/f_H=1$.\ Thus, the total solar capture rate is given by $C\hsm =\hsm C_H^{}\!+\!C_L^{}$.

\vs 

Then, we further make two assumptions that maximize the observable neutrino signal: 
the perfect capture-annihilation equilibrium, $\Gamma_{\rm ann}\!=\!C/2\,$, 
and the $100\%$ annihilation branching fraction 
into $W^+W^-$.\ 
For this, we define a ratio,  
\begin{equation}
{\cal R}_\odot^{} = 
\frac{C/2}{~\Gamma_{\rm lim}^{WW}(m_{\chi}^{})~}, 
\label{eq:ratio}
\end{equation}
where $\Gamma_{\rm lim}^{WW}$ is the 
iDM annihilation rate into $W^+W^-$ 
as constrained by the IceCube solar neutrino flux\,\cite{PospelovRamani2026}, 
which is a function of $m_{\chi}^{}$.\ 
In Fig.\,\ref{fig:solar}, we plot ${\cal R}_\odot^{}$ as a function of iDM mass.\ 
We choose $\mu_{N}^{}\Delta m/m_{N}\!\sim\! 248\,$keV for $\Delta m$, and set the DM-nucleon scattering cross section to be the best fit of the LZ event excess.\ 
As can be seen, the contribution of the heavier state is dominant, and the lighter state gives only subleading contribution 
in the higher-mass region. 

\vs

In the LZ-relevant region, 
the total contribution from both lighter and heavier states
remains comfortably below the IceCube-equivalent 
exclusion bound at 90\%\,C.L. (the red line) 
even under all the above overestimates of solar neutrino signals.\  
Hence, as is clear, already at the level of the 
initial capture rate, 
the LZ-compatible exothermic case predicts much less 
solar neutrino signals than the near-threshold endothermic case 
(such as Higgsino models), 
due to the smaller cross section in the exothermic case 
required by fitting the LZ data.\ 
Hence, a broad exothermic parameter space remains viable 
even under the assumptions deliberately chosen to overestimate 
the neutrino signals for the IceCube detection.


\vspace*{-2mm}
\section{\hspace*{-3mm}Conclusions}
\label{sec:5}
\label{sec:conclusions}
\vspace*{-1mm}

Inelastic DM (iDM) scattering produces a characteristic 
nuclear recoil (NR) energy that differs from the elastic case.\ 
For both the exothermic and endothermic iDM cases, 
the incident speed required for a low-energy recoil diverges.\  
This feature permits high-energy NR events 
without the low-energy spectral concentration of 
conventional elastic, spin-independent WIMP scattering.\ 
Mass-splittings of several hundred keV are relevant to the LZ candidate event 
for weak-scale and heavier DM particle. 

\vs 

In this work, we performed a model-independent study 
for both the exothermic and endothermic iDM and their NR signals 
in the LZ experiment.\  
In Sec.\,\ref{sec:2}, we analyzed the kinematics of the inelastic DM scattering 
and computed the corresponding NR spectra, 
showing that the high-energy NR event excess can be naturally explained 
by the iDM.\  
In Sec.\,\ref{sec:3}, we scanned over the iDM parameter space of 
$10\GeV\!\!\leqq\!\! m_\chi\!\!\leqq\!\!1000\GeV$ 
and $-500\keV\!\!\leqq\!\Delta m\!\leqq\!1000\keV$, 
and derived the best fits and 90\% C.L. intervals on the DM-proton cross section
as shown in Fig.\,\ref{fig:2} and Fig.\,\ref{fig:3}.\  
We further derived the bounds in 
the $(m_\chi,\Delta m)$ plane at 68\%\,C.L. and 90\%\,C.L.\  
as presented in Fig.\,\ref{fig:4}.\ 
Our results show that the LZ event favors the DM mass-splitting ranges within  
$165\,{\rm keV}\!<\!|\Delta m|\!<\!350\,{\rm keV}$ for the endothermic iDM 
and within $165\,{\rm keV}\!<\!\Delta m\!<\!1000\,{\rm keV}$ for the exothermic iDM.\  
Also, the exothermic case can account for the LZ data with 
much smaller DM-nucleon cross sections.\ 
The abundance, longevity and solar capture effects of the exothermic iDM 
were discussed in Sec.\,\ref{sec:4}.\  
We demonstrated that it is feasible for the heavier iDM state 
to be stable at the cosmological time scale, 
and the neutrino flux from the annihilation of solar captured exothermic iDM
is well below the IceCube limits because of the much smaller DM-nucleon cross sections. 

\vspace*{2mm}
\noindent 
{\bf Acknowledgments}
\\[1mm]
The works of ZTF, HJH and YCW were supported in part
by the National Natural Science Foundation of China (NSFC)
(Grant Nos.\,12435005 and 12175136), by Shenzhen Science and Technology Program
(Grant No.\,JCYJ2024 0813150911015), by the State Key Laboratory of Dark Matter Physics,
by the Key Laboratory for Particle Astrophysics and Cosmology (MOE), 
and by the Shanghai Key Laboratory for Particle Physics and Cosmology. 

\end{document}